\documentclass[12pt,a4paper]{article}
\usepackage{a4wide}
\usepackage{graphicx}
\usepackage{hyperref}

\begin{document}

\begin{titlepage}
\begin{flushright}
LU TP 14-41\\
December 2014\\
\end{flushright}
\vfill
\begin{center}
{\Large\textbf {CHIRON}: a package for ChPT numerical results
at two loops$^\dagger$}
\vfill
{\bf Johan Bijnens}\\[0.3cm]
{Department of Astronomy and Theoretical Physics, Lund University,\\
S\"olvegatan 14A, SE 223-62 Lund, Sweden}
\end{center}
\vfill
\begin{abstract}
This document describes the package {\textsc CHIRON}
which includes two libraries, \texttt{chiron} itself and \texttt{jbnumlib}.

\texttt{chiron} is a set of routines useful for two-loop numerical results in
Chiral Perturbation Theory (ChPT). 
It includes programs for the needed
one- and two-loop integrals as well as
routines to deal with the ChPT parameters.
The present version includes everything needed for the masses, decay
constants and quark-antiquark vacuum-expectation-values.
An added routine calculates consistent values for the masses and decay constants
when the pion and kaon masses are varied.
In addition a number of finite volume results are included:
one-loop tadpole integrals, two-loop sunset integrals
and the results for masses and decay constants.

The numerical routine library \texttt{jbnumlib}
contains the numerical routines used in \texttt{chiron}. Many are to a large
extent simple {\textsc C++} versions of routines in the CERNLIB numerical
library. Notable exceptions are the dilogarithm and
the Jacobi theta function implementations.

This paper describes what is included in \textsc{CHIRON} v0.50.
\end{abstract}
\vfill
\end{titlepage}

\tableofcontents

\section{Introduction}

Chiral Perturbation Theory (ChPT)
is the low-energy effective field theory of QCD.
It was introduced by Weinberg, Gasser and Leutwyler
\cite{Weinberg:1978kz,Gasser:1983yg,Gasser:1984gg} and the present state of
the art are calculations performed at two-loop level.
A review is \cite{Bijnens:2006zp} but many more exists. The long term goal
of this project is to make available all these calculations with a consistent
interface in {\textsc C++}. Many of the original programs were written in
{\textsc FORTRAN77} and are available on request from the authors, but they are
not always consistent in the interfaces and the use of common blocks
for moving parameters around has occasionally lead to difficult to find errors.

A general knowledge of {\textsc C++} is assumed throughout this paper.
The routines are at present not guaranteed to be thread-safe,
some global variables inside the various files are used
for the loop functions and integration routines. These are however never used
for setting outside the files, there are always functions provided for this.
It is recommended to always use these.
The routines return double precision types if not indicated directly.

Kheiron, $X\varepsilon\iota\rho\omega\nu$, or Chiron, was the eldest and
wisest of the Centaurs, half-horse men of Greek mythology.
His name is derived from the Greek word for hand (Kheir)
which also formed the basis of the word chiral which is why his name was chosen
for this package \cite{chiron}.

The license chosen is the General Public License v2 or later from
the Free Software Foundation \cite{GPLv2}.

The \texttt{chiron} routines have mainly been tested against
the \textsc{FORTRAN}
codes of the original publications. These were in turn implemented
in at least two independent versions originally. 
The \texttt{jbnumlib}
routines have their output compared with the original \textsc{CERNLIB}
routines in case they were simple translations to \textsc{C++}.
In the other cases, the
tests are described in the relevant sections.

This paper describes what is included in \textsc{CHIRON} v0.50.
The package itself is available from \cite{chironsite}.
The included files and how to install it is described in Sect.~\ref{setup}.
The numerical routines included in \texttt{jbnumlib} are described
in Sect.~\ref{jbnumlib}. Some short comments on
ChPT notation are in Sect.~\ref{chptnotation}. The main part
describing the contents of \texttt{chiron} are Sects.~\ref{datastructures}
to \ref{massdecayvevV}. Sect.~\ref{datastructures} contains the objects
implemented to deal with input data for the ChPT calculations.
A large part of the work is in implementing the relavant loop integrals,
especially those at finite volume. This is the content of
Sect.~\ref{loopintegrals}. The simplest quantities
are masses, decay constants and vacuum-expectation-values. 
The functions for these are discussed in Sect.~\ref{massdecayvev}
and the finite volume extensions for masses and decay constants
in Sect.~\ref{massdecayvevV}. Some comments about errors, some warnings
about the use of the routines and definitions in ChPT as well as a number
of planned/possible extensions are discussed in Sect.~\ref{various}.
A short summary is given in the final section.

\section{Files and setup}
\label{setup}

The package is delivered as a gzipped tarred file (\texttt{chironvvvv.tar.gz}),
where
vvvv is version information. Untarring it creates a directory
\texttt{chironvvvv}, which is referred to as the root directory below.
  
The package has a number of subdirectories when delivered.
The root directory contains a Makefile and files
COPYING, INSTRUCTION and GUIDELINES. The main instructions to
produce {\textsc CHIRON} are to do first ``make libjbnumlib.a'' to
produce the numerical library and then ``make libchiron.a''
to produce the main library or simply ``make'' to do both.
The latter also puts the newly produced library versions in the subdirectory
\texttt{lib}.
To install, put the the two library files
and the content of the \texttt{include} directory somewhere where they
can be linked to and included. Linking should be indicated on the compile line
with the options ``-lchiron -ljbnumlib''.

The subdirectory \texttt{doc} contains this manuscript, a filelist and
possibly more files in future versions. The subdirectory \texttt{src}
contains the source files
and \texttt{include} the various header files. The subdirectory \texttt{test}
contains a number of testing programs where the names ``testyyy.cc'' indicates
a program testing the code in the source file ``yyy.cc''. A testing program
can be compiled using ``make testyyy'' in the root directory. The program
``a.out'' should then produce the output as shown in the file ``testyyy.dat''
in the subdirectory \texttt{testoutputs}.
The test subdirectory contains in addition the file LiCiBE14.dat
with the latest determination of the LECs \cite{Bijnens:2014lea}.

\section{\texttt{jbnumlib}}
\label{jbnumlib}

The functions in this section are included to make the program collection
self-contained. They are mainly implementations of well known programs in $C++$
and in particular many of the routines are a port to $C++$
from the {\textsc CERNLIB} \cite{cernlib} {\textsc FORTRAN} routines.
Some, as mentioned in the respective texts, are fully original.
The definitions are all in \texttt{jbnumlib.h} and contained in
\texttt{libjbnumlib.a}. The implementations are in the files
menstioned in each subsection below.
In order to avoid conflict with other implementations all routines
in this section have names starting with \texttt{jb}.
The exact interface is best checked by looking in the include file
\texttt{jbnumlib.h}.

\subsection{Special functions}

\subsubsection{Dilogarithm or $\mathrm{Li}_2(x)$: \texttt{jbdli2(x)}}

The way the vertex integrals are implemented requires a Spence or
$\mathrm{Li}_2$
function which returns complex values for all possible complex inputs.
The routine implemented uses the algorithm given in \cite{'tHooft:1978xw},
Appendix A
up to Bernouilly number $B_{28}$. Defined in \texttt{jbdli2.cc}.
For real numbers the output has been compared to that of the
\textsc{CERNLIB} routine \texttt{DDILOG}.
It has also been checked that the function satisfies a number of the
relations between values with different arguments that were not used in
its evaluation.

\subsubsection{Bessel functions}

The modified Bessel functions $I_0,I_1,K_0,K_1$ with real
arguments are available as \texttt{jbdbesi0}, \texttt{jbdbesi1}, 
 \texttt{jbdbesk0} and \texttt{jbdbesk1}. These are implemented in
\texttt{jbdbesio} which is a simple port to \textsc{C++} of the
\texttt{CERNLIB} routines \texttt{dbesi0},\ldots.
In addition the modified Bessel functions $K_1,K_2,K_3$
are available as  \texttt{jbdbesk2},  \texttt{jbdbesk3}
and \texttt{jbdbesk4}. These are evaluated using the recursion relations
from $K_0$ and $K_1$.

\subsubsection{Theta functions}

The functions defined are related to the Jacobi theta functions.
\texttt{jbdtheta30(q)} returns the function
\begin{equation}
\label{deftheta30}
\theta_{30}(q) = 1+2\sum_{n=1,\infty} q^{(n^2)}\,.
\end{equation}
It uses the idea behind the \texttt{CERNLIB} routines \texttt{DTHETA}.
For small $q$ it simply sums the series (\ref{deftheta30}) and for
larger $q$ it uses the modular invariance and a series in the changed variable
instead. The accuracy has been checked by running both series too much higher
orders and comparing the two results. Implemented in \texttt{jbdtheta30.cc}.
A function without the 1 which is needed to keep accuracy for small
$q$ is available as \texttt{jbdtheta30m1}. 
Implemented in \texttt{jbdtheta30m1.cc}.

\texttt{jbdtheta32(q)} returns the function
\begin{equation}
\label{deftheta32}
\theta_{32}(q) = 2\sum_{n=1,\infty}n^2 q^{(n^2)}\,.
\end{equation}
For small $q$ it simply sums the series (\ref{deftheta32}) and for
larger $q$ it uses the modular invariance and the derivative
of the series in the changed variable
The accuracy has been checked by running both series too much higher
orders and comparing the two results. Defined in \texttt{jbdtheta32.cc}.
A function with $n^4$ multiplying $q^{(n^2)}$ is available
as \texttt{jbdtheta34} and implemented in \texttt{jbdtheta34.cc}.

\subsubsection{Higher dimensional theta functions}
 
There are higher dimensional generalizations of the theta functions.
These satisfy a more general modular invariance which can be used
to get much faster convergence series. The functions defined below use
some of these possible optimizations.

The basic function is the 2-dimensional generalization
\begin{equation}
\theta^{(2d)}_{0}(\alpha,\beta,\gamma) = \sum_{n_1,n_2=-\infty,\infty}
e^{-\alpha n_1^2-\beta n_2^2-\gamma (n_1-n_2)^2}\,.
\end{equation}
This function is implemented as \texttt{jbdtheta2d0} with arguments
$\alpha,\beta,\gamma$  in \texttt{jbdtheta2d0.cc}.
In addition the function with the one removed is also
available as \texttt{jbdtheta2d0m1}
implemented in \texttt{jbdtheta2d0m1.cc}.

The function with the exponential multiplied by $n_1^2$ is
called \texttt{jbdtheta2d02} and implemented in \texttt{jbdtheta2d02.cc}.

\subsection{Integration routines}

An adaptive gaussian quadrature routine \texttt{jbdgauss} and an adaptive
integration routine \texttt{jbdcauch}, that integrates symmetrically
around a singularity, are included.
These are ports of the {\textsc CERNLIB} routines
\texttt{dgauss} and \texttt{dcauch} respectively. Implementation is in
\texttt{jbdgauss.cc} and \texttt{jbdcauch.cc}. The complex equivalent is
\texttt{jbwgauss} in \texttt{jbwgauss.cc}.

The higher dimensional integration \textsc{CERNLIB} routine \texttt{DADMUL}
based on \cite{radmulpaper} has been ported to \texttt{C++}
and a simple interface for two and three-dimensional integration
implemented as \texttt{jbdad2} and \texttt{jbdad3}.
The code can be found in \texttt{jbdadmul.cc}.

\section{ChPT notation}
\label{chptnotation}

The notation used in ChPT is not fully unique. The notation used
in this program collection is the main one used by the author
and his collaborators. The main point to be observed is that \texttt{chiron}
uses a normalization for the decay constants with $F_\pi\approx92~$MeV.

For the low-energy-constants, we use the conventions of \cite{Gasser:1984gg}
and \cite{Bijnens:1999sh,Bijnens:1999hw} with dimensionless
renormalized couplings $L_i^r$ and $C_i^r$.

The lowest order couplings are denoted by $F_0$ and $B_0$. The quark masses
are $\hat m=m_u=m_d$, note that we work in the isospin limit, and $m_s$.
The lowest order masses\footnote{Note that the programs use an internal
convention where \texttt{mhat}$=2B_0\hat m$ and \texttt{mstrange}$=2B_0 m_s$.} 
 are given by
\begin{equation}
\label{lomasses}
m_{\pi\,0}^2 = 2B_0\hat m\,,
\quad
m_{K\,0}^2 = B_0\left(\hat m+m_s\right)\,,\quad
m_{\eta\,0}^2 = \frac{B_0}{3}\left(2\hat m+4m_s\right)\,.
\end{equation}

\section{Data structures}
\label{datastructures}

This section discusses the structures available for dealing with masses,
$F_\pi$, the $L_i^r$ and $C_i^r$, and the subtraction constant $\mu$.
Note that $\mu$ is present in all three data structures and the user should
make sure that their use is consistent.
The default-value mechanism in \texttt{C++} has been used to define the values
when the constructors are called with less than the full data needed.

These data structures are implemented as classes.

Note that we assume dimensional
units in GeV, but if all dimensional inputs are scaled accordingly
the routines give the correct answer. However, typical precisions set
are assuming ChPT applications with dimensional units in powers of GeV.

\subsection{physmass: Masses, $F_\pi$, $\mu$}

The physical masses are defined in a class \texttt{physmass} defined in
 \texttt{inputs.h} and implemented in \texttt{inputs.cc}. 

Private data members: \texttt{mpi, mk, meta, fpi, mu}.

Typical declaration: \texttt{physmass mass1(0.135,0.495,0.548,0.0922,0.77);}
The numbers given above are also the defaults. 

These are the physical pion, kaon and eta mass, $m_\pi,m_K,m_\eta$
the physical pion decay constant, $F_\pi$ and the subtraction point $\mu$.
The default constructor puts them all at some reasonable 
values but they can be created with any
number of the inputs specified, starting from the left.
In addition there are functions \texttt{void setmpi(double)} etc., defined
that set one of the values only. These use the same default values.

The values can be obtained from the function
\texttt{void out( mpi, mk, meta, fpi, mu)} that returns all of them via
referenced doubles. Functions \texttt{double getmpi(void)} etc. are defined
that return one of the values.

The output/input stream format is defined as well so \texttt{cout << mass1}
and \texttt{cin >> mass1} make sense. 
The input stream should have the same format as the output stream produces.
This works for all streams, not just the standard \texttt{cout} and
\texttt{cin}.

A test for equality is defined which checks that all data members agree to 7 
significant digits. So expressions like \texttt{if(mass1 == mass2)} can be used.
This is relative precision, so it is assumed no mass is zero here.
The reason for not using exact equality is that calculated masses might not be
exactly the same using double precision variables. 

\subsection{Classes for the NLO LECS: \texttt{Li}}

The class for dealing with the next-to-leading-order (NLO)
low-energy constants (LECs) defined in \cite{Gasser:1984gg} is
named \texttt{Li}.
This is implemented in \texttt{Li.cc} and defined in \texttt{Li.h}.

The \texttt{Li} class has 13 double precision variables to store the
LECs $L_1^r,\ldots,L_{10}^r,H_1^r,H_2^r$ and the subtraction scale $\mu$.
It also contains a string with a name for the set of constants.
The LECs default to zero, the scale to 0.77 and the name to ``nameless Li.''
When the LECs are referred to with numbers, 11,12 correspond to $H_1^r,H_2^r$
respectively.

Typical declarations are:\\
\texttt{Li Li1;}
\texttt{Li Lifitall(0.88e-3,0.61e-3,-3.04e-3,0.75e-3,0.58e-3,0.29e-3,}\\
\texttt{-0.11e-3,\allowbreak-0.18e-3,5.93e-3,0.,0.,0.,0.77,"fit All");}

Operations defined on the \texttt{Li}: overloaded operators are defined
such that sets of \texttt{Li} can be added or subtracted and multiplied by
a number (double).The output/input stream format is defined as well so
\texttt{cout $<<$ Li1}
and \texttt{cin $>>$ Li1} make sense. The input stream should have the same
format as the output stream produces.

Member functions that can be used to set the parameters
are \texttt{setli} which takes an integer and a double (in either order) as
argument to set the corresponding LEC to the double, \texttt{setmu} which
sets the scale\footnote{This sets the scale simply, it does \emph{not} change
the numerical values of the LECs.}
 and \texttt{setname} which changes the name of the set of $L_i^r$.

Output member functions exist to obtain a single LEC, \texttt{out(int)},
or the 10 $L_i^r$, the 10 $L_i^r$ and $\mu$, the 12 LECs,
the 12 LECs and $\mu$, and the 12 LECs, $\mu$ and the name.
These are all called out and return the results via a reference to
10, 11, 12, 13 double or 13 double and a string variable.

The member function \texttt{changescale} changes the scale $\mu$ and changes
the $L_i^r$ and $H_i^r$ according the scale dependence as obtained first
in \cite{Gasser:1984gg}.

There are also three functions defined that return a set of random NLO LECs.
These are \texttt{Lirandom} wich gives each LEC a random value between
$\pm1/(16\pi^2)$. \texttt{LirandomlargeNc} does the same but leaves $L_4^r$,
$L_6^r$ and $L_7^r$ zero. Finally, \texttt{LirandomlargeNc2} does the same but 
$L_4^r$, $L_6^r$ and $L_7^r$ get a random value between $\pm(1/3)/(16\pi^2)$.
Note that $1/(16\pi^2)\approx 0.0063$ so the ranges include the values of the
fitted $L_i^r$. The random numbers are generated using the system
generator \texttt{rand()} so we recommend initializing using something like
\texttt{srand(time (0))}. These latter functions were used in the
random walks in the $L_i^r$ in \cite{Bijnens:2011tb}.

\subsection{Classes for the NNLO LECS: \texttt{Ci}}

The class for dealing with the next-to-next-to-leading-order (NNLO)
low-energy constants (LECs) defined in \cite{Bijnens:1999sh} is
named \texttt{Ci}.
This is implemented in \texttt{Ci.cc} and defined in \texttt{Ci.h}.
Note that this set of routines uses the convention where the $C_i^r$
are dimensionless. The parameters in the Lagrangian have dimension
mass$^{(-2)}$ but the definition of the subtracted $C_i^r$ in
\cite{Bijnens:1999hw} is dimensionless.
Going from one-convention to the other is with the appropriate
power\footnote{The definition is with the chiral limit value $F_0$
but the difference is higher order.}
of $F_\pi$.

The \texttt{Ci} class has as private members
a double precision array \texttt{Cr[95]}, which holds the $C_i^r$, $i=1,94$
in \texttt{Cr[i]}, the scale mu and string for the name.
Defaults are zero for all the $C_i^r$, 0.77~GeV for the scale
and ``nameless Ci'' for the name.

Constructors are provided with as input a double array Cr[95],
the scale and a name or a scale and a name or a scale only or no input.
Typical declarations are:\\
\texttt{Ci Ci1,Ci2(1.0),Ci3(Crr,0.8,"a nice set")} where \texttt{Crr}
is defined as \texttt{double Crr[95]}.

An additional constructor is provided that has as input the resonance parameters
where the resonance model is the simple version described
in Sect.~5 of \cite{Amoros:2000mc}.

Operations defined on the \texttt{Ci} are: overloaded operators are defined
such that sets of \texttt{Ci} can be added or subtracted and multiplied by
a number (double).The output/input stream format is defined as well so
\texttt{cout $<<$ Ci1}
and \texttt{cin $>>$ Ci1} make sense. The input stream should have the same
format as the output stream produces.

Member functions that can be used to set the parameters
are \texttt{setci} which takes an integer and a double (in either order) as
argument to set the corresponding LEC to the double, \texttt{setmu} which
sets the scale\footnote{This sets the scale simply, it does \emph{not} change
the numerical values of the LECs.}
 and \texttt{setname} which changes the name of the set of $C_i^r$.

Output member functions exist to obtain a single LEC, \texttt{out(int)},
or the \texttt{Cr[95]}, or \texttt{Cr[95]} and the scale, or  \texttt{Cr[95]},
scale and name.
These are all called out and return the results via references
to the string and scale and the array.

The member functions \texttt{changescale(double,Li)}
and \texttt{changescale(Li,double)} change the scale $\mu$ and changes
the $L_i^r$, $H_i^r$ and $C_i^r$ according the scale dependence as obtained
first in \cite{Bijnens:1999hw}. Note that the \texttt{Li} set here has also the
scale and values changed accordingly, not only the $C_i^r$.

There are also three functions defined that return a set of random NLO LECs.
These are \texttt{Cirandom} which gives each LEC a random value between
$\pm1/(16\pi^2)$. \texttt{CirandomlargeNc} does the same but leaves
the large-$N_c$ suppressed constants zero.
 Finally, \texttt{CirandomlargeNc2} does the same but the large-$N_c$
suppressed constants get a value between $\pm(1/3)/(16\pi^2)$.
The random numbers are generated using the system
generator \texttt{rand()} so we recommend initializing using something like
\texttt{srand(time (0))}. Typically these values of the $C_i^r$ are somewhat
on the large side when fitting data.

\section{Loop integrals}
\label{loopintegrals}

Most of the integrals used have been treated in many places.
I refer only to the papers where our particular notation has been defined
and/or the method used to evaluate them was developed.
When comparing with other packages, keep in mind the differences in
subtraction and/or differences in defining the integrals.

\subsection{One-loop integrals}
\label{oneloop}

\subsubsection{Tadpoles}
\label{tadpoles}

These integrals have been defined in \cite{Amoros:1999dp} and correspond to
the finite parts of the integral
\begin{equation}
A(n,m^2) = \frac{1}{i}\int\,\frac{d^dp}{(2\pi)^d}\frac{1}{(p^2-m^2)^n}\,.
\end{equation}
After the subtraction and renormalization as usual in ChPT, we are left
with the finite four-dimensional part $\bar A(n,m^2)$ which is implemented
as \texttt{Ab(n,msq,mu2)} and the $n=1,2,3$ as \texttt{Ab,Bb,Cb} respectively
with arguments \texttt{msq,mu2}.
These functions are defined in \texttt{oneloopintegrals.h}
and implemented in \texttt{oneloopintegrals.cc}. 

\subsubsection{Bubble integrals}
\label{bubble}

These have been defined in \cite{Amoros:1999dp,Bijnens:2002hp}
\begin{eqnarray}
B(m_1^2,m_2^2,p^2) &=& \frac{1}{i}\int 
\frac{d^dq}{(2\pi)^d} \frac{1}{(q^2-m_1^2)((q-p)^2-m_2^2)}\,,\nonumber\\
B_\mu(m_1^2,m_2^2,p^2) &=& \frac{1}{i}\int 
\frac{d^dq}{(2\pi)^d} \frac{q_\mu}{(q^2-m_1^2)((q-p)^2-m_2^2)}\nonumber\\
&=& p_\mu B_1(m_1^2,m_2^2,p^2)\,,\nonumber\\
B_{\mu\nu}(m_1^2,m_2^2,p^2) &=& \frac{1}{i}\int 
\frac{d^dq}{(2\pi)^d} \frac{q_\mu q_\nu}{(q^2-m_1^2)((q-p)^2-m_2^2)}\nonumber\\
&=& p_\mu p_\nu B_{21}(m_1^2,m_2^2,p^2)+g_{\mu \nu} B_{22}(m_1^2,m_2^2,p^2)\,,\nonumber\\
B_{\mu\nu\alpha}(m_1^2,m_2^2,p^2) &=& \frac{1}{i}\int 
\frac{d^dq}{(2\pi)^d} \frac{q_\mu q_\nu q_\alpha}{(q^2-m_1^2)((q-p)^2-m_2^2)}\nonumber\\
&& \hspace{-2cm}
= p_\mu p_\nu p_\alpha B_{31}(m_1^2,m_2^2,p^2) + 
(p_\mu g_{\nu \alpha} + p_\nu g_{\mu \alpha} + p_\alpha g_{\mu \nu})B_{32}(m_1^2,m_2^2,p^2)
\,.
\end{eqnarray}
Again one needs to do the subtraction and renormalization with ChPT convention.
The analytical values can be obtained using the methods
of \cite{'tHooft:1978xw} for the $B$ integral and the others can be reduced
to it using the methods of \cite{Passarino:1978jh}.
All functions have been implemented via a method that does the integration
over the Feynman parameter $x$ numerically.
These have real arguments \texttt{m1sq,m2sq,psq,mu2} and are called
\texttt{Bbnum,B1bnum,B21bnum,B22bnum,}\\ \texttt{B31bnum, B32bnum} and return a complex
value. The analytical evaluation has been implemented
in \texttt{Bb,B1b,B21b,B22b} with the same arguments and a complex
return value. The simpler analytical expression for the case of the
two masses equal has been implemented analytically for \texttt{Bb} and
\texttt{B22b} called
with argument \texttt{msq,psq,mu2}.

For the cases with numerical integrations, the precision can be set
using\\\texttt{setprecisiononeloopintegrals(double)} and
obtained by\\ \texttt{getprecisiononeloopintegrals(void)}.

All functions are defined in \texttt{oneloopintegrals.h}
and implemented in\\ \texttt{oneloopintegrals.cc}. 

\subsection{Sunset integrals}
\label{sunset}

These give the sunsetintegral loop integral functions
$H^F,H_{1}^F, H_{21}^F$
defined in App.  of \cite{Amoros:1999dp}.
The definition in finite volume is
given in (\ref{defsunset}) below.
$H_{31}^F$ is the function
multiplying the $p_\mu p_\nu p_\rho$ part of the integral with
$r_\mu r_\nu r_\rho$. These exist in a real version valid below threshold and
a complex version valid everywhere. The method used is derived
in \cite{Amoros:1999dp}. 

The derivative w.r.t. $p^2$ is included for the real version of
$\overline H,\overline H_{1}, \overline H_{21}$. Functions defined
in \texttt{sunsetintegrals.h} and implemented in \texttt{sunsetintegrals.cc}.

Input arguments are real and are $m_1^2, m_2^2,m_3^2,p^2,\mu^2$.
Naming conventions are \texttt{hh,hh1, hh21,hh31} for the real versions valid
below threshold and the complex versions valid
everywhere \texttt{zhh,zhh1,zhh21,zhh31}. The real versions are normally faster
when applicable. In addition, wave-function renormalization requires
some derivatives w.r.t. the external momentum $p^2$.
These are encoded in \texttt{hhd,hh1d,hh21d} and \texttt{zhhd,zhh1d,zhh21d}
for the real and complex case respectively.

The precision of the numerical integrations can be set
using\\ \texttt{setprecisionsunsetintegrals(double)} and
obtained by\\ \texttt{getprecisionsunsetintegrals(void)}.

These functions are defined in \texttt{sunsetintegrals.h}
and implemented in\\ \texttt{sunsetintegrals.cc}.
 
\subsection{One-loop finite volume integrals}

The methods used for these are derived in detail in \cite{Bijnens:2013doa},
references to earlier literature can be found there.
The integrals used here are given in the Minkowski conventions as
defined in \cite{Bijnens:2014dea}.
All of the integrals are available with two different methods, one using
a summation over Bessel function and the other an integral over a
Jacobi theta function. The versions included at present are using
periodic boundary conditions, all three spatial sizes of the same length $L$
and the time direction of infinite extent.

\subsubsection{Tadpoles}

The tadpole integrals $A$ and $A_{\mu\nu}$ are defined as
\begin{equation}
\left\{A(m^2),A_{\mu\nu}(m^2)\right\}
 = \frac{1}{i}\int_V\frac{d^d r}{(2\pi)^d}
\frac{\left\{1,r_\mu r_\nu\right\}}{(r^2-m^2)}\,.
\end{equation}
The $B$ tadpole integrals are the same but with a doubled propagator.

The subscript $V$ on the integral indicates that the integral is a discrete
sum over the three spatial components and an integral over the remainder. 
At finite volume, there are more Lorentz-structures possible. 
The tensor $t_{\mu\nu}$, the spatial part of the Minkowski metric
$g_{\mu\nu}$, is needed for these.
The functions for $A_{\mu\nu}$ are
\begin{equation}
A_{\mu\nu}(m^2) = g_{\mu\nu}A_{22}(m^2)+t_{\mu\nu}A_{23}(m^2)\,.
\end{equation}
In infinite volume $A_{22}$ is related to $A$ and $A_{23}$ vanishes.
We denote the finite volume part by a superscript $V$ and
one should remember that for the full integrals, the infinite volume results
of Sect.~\ref{tadpoles} need to be added.

The functions are defined as \texttt{AbVt(msq,L)},  \texttt{BbVt(msq,L)},
 \texttt{AbVb(msq,L)}, \texttt{BbVb(msq,L)}. The last letter
indicates whether they are computed with the theta function or Bessel
function method.
The results were checked by comparing against each other and
by comparing with the independent Bessel function implementation done in
\cite{Bijnens:2006ve}.

The functions
\texttt{A22bVt(msq,L)},  \texttt{A22bVt(msq,L)},
and \texttt{A23bVb(msq,L)}, \texttt{A23bVb(msq,L)} are available as well.

\texttt{setprecisionfinitevolumeoneloopt(Abacc,Bbacc,printout)}
and\\ \texttt{setprecisionfinitevolumeoneloopt(maxsum,Bbacc,printout)}
set the precision.
The last variable printout is a logical variable which can be set to
true or false, default is false. The first and second argument give the
(mainly absolute) precision of the numerical integration for the
tadpole and bubble integral numerical integrations.
maxsum indicates how far the sum over Bessel function is taken.
Maximum at present is 400.

These functions are defined in \texttt{finitevolumeoneloopintegrals.h}
and implemented in \texttt{finitevolumeoneloopintegrals.cc}.

\subsubsection{Bubble integrals}

Not implemented in this version.

\subsection{Sunset finite volume integrals}

Sunset integrals are defined as
\begin{equation}
\label{defsunset}
\left\{H,H_\mu,H_{\mu\nu}\right\} =
\frac{1}{i^2}\int_V\frac{d^d r}{(2\pi)^d}\frac{d^d 1}{(2\pi)^d}
\frac{\left\{1,r_\mu,r_\mu r_\nu\right\}}
{\left(r^2-m_1^2\right)\left(s^2-m_2^2\right)\left((r+s-p)^2-m_3^2\right)}\,.
\end{equation}
The subscript $V$ indicates that the spatial dimensions are a discrete
sum rather than an integral.
The conventions correspond to those in infinite volume of
\cite{Amoros:1999dp}.
Integrals with the other momentum $s$ in the numerator are related
using the trick shown in \cite{Amoros:1999dp} which remains
valid at finite volume in the cms frame \cite{Bijnens:2013doa}.

In the cms frame we define the functions\footnote{In the cms frame
$t_{\mu\nu}= g_{\mu\nu}-p_\mu p_\nu/p^2$ but the given separation
appears naturally in the calculation \cite{Bijnens:2013doa}. It also avoids
singularities in the limit $p\to0$.}
\begin{eqnarray}
\label{defHi}
H_\mu &=& p_\mu H_1\,
\\\nonumber
H_{\mu\nu} &=& p_\mu p_\nu H_{21} + g_{\mu\nu} H_{22} + t_{\mu\nu} H_{27}\,.
\end{eqnarray}
The arguments of all functions in the cms frame are
$(m_1^2,m_2^2,m_3^2,p^2)$.
These functions satisfy the relations, valid in finite
volume \cite{Bijnens:2013doa},
\begin{eqnarray}
\label{Hrelations}
H_1(m_1^2,m_2^2,m_3^2,p^2)+
H_1(m_2^2,m_3^2,m_1^2,p^2)+
H_1(m_3^2,m_1^2,m_2^2,p^2)
&=& H(m_1^2,m_2^2,m_3^2,p^2)\,,
\nonumber\\
p^2 H_{21}+d H_{22} + 3 H_{27}-m_1^2 H &=& A(m_2^2)A(m_3^2)\,.
\end{eqnarray}
The arguments of the sunset functions in the second relation are
all $(m_1^2,m_2^2,m_3^2,p^2,L,\mu^2)$. ($L$ only for the finite volume part).

We split the functions in an infinite volume part, $\tilde H_i$, and a finite
volume correction, $\tilde H^V_i$, with
$H_i=\tilde H_i+\tilde H^V_i$. The infinite volume part has been
discussed above.
For the finite volume parts we define
\begin{eqnarray}
\tilde H^V &=& \frac{\lambda_0}{16\pi^2}
 \left(A^V(m_1^2)+A^V(m_2^2)+A^V(m_3^2)\right)
 +\frac{1}{16\pi^2}
    \left(A^{V\epsilon}(m_1^2)+A^{V\epsilon}(m_2^2)+A^{V\epsilon}(m_3^2)\right)
\nonumber\\&&
 +H^V\,,
\nonumber\\
\tilde H^V_1 &=& \frac{\lambda_0}{16\pi^2}\frac{1}{2}
 \left(A^V(m_2^2)+A^V(m_3^2)\right)
 +\frac{1}{16\pi^2}\frac{1}{2}
    \left(A^{V\epsilon}(m_2^2)+A^{V\epsilon}(m_3^2)\right)
 +H^V_1\,,
\nonumber\\
\tilde H^V_{21} &=& \frac{\lambda_0}{16\pi^2}\frac{1}{3}
 \left(A^V(m_2^2)+A^V(m_3^2)\right)
 +\frac{1}{16\pi^2}\frac{1}{3}
    \left(A^{V\epsilon}(m_2^2)+A^{V\epsilon}(m_3^2)\right)
 +H^V_{21}\,,
\nonumber\\
\tilde H^V_{27} &=& \frac{\lambda_0}{16\pi^2}
 \left(A^V_{23}(m_1^2)+\frac{1}{3}A_{23}(m_2^2))+\frac{1}{3}A^V_{23}(m_3^2)\right)
\nonumber\\&&
 +\frac{1}{16\pi^2}
    \left(A^{V\epsilon}_{23}(m_1^2)
+\frac{1}{3}A^{V\epsilon}_{23}(m_2^2+\frac{1}{3}A^{V\epsilon}_{23}(m_3^2))\right)
 +H^V_{27}\,.
\end{eqnarray}
The finite parts are defined differently from the
infinite volume case in \cite{Amoros:1999dp}.
The parts with $A^{V\epsilon}$ are removed here as well.

The functions $H^V_i$ can be computed with the methods
of \cite{Bijnens:2013doa}.
They correspond to adding the parts labeled with $G$ and $H$ in Sect. 4.3 and
the part of Sect. 4.4 in \cite{Bijnens:2013doa}.

They are implemented as functions \texttt{hhVt,hh1Vt,hh21Vt,hh22Vt,hh27Vt}
with arguments \texttt{m1sq,m2sq,m3sq,psq,L,mu2}.
The derivatives w.r.t. $p^2$ exist as
\texttt{hhdVt,hh1dVt,\\hh21dVt,hh22dVt,hh27dVt}.
These are the functions using the theta function method. Those using
the Bessel function method are implemented with a b instead of t as last letter
in the name. The arguments are the same.

For all cases discussed we have done checks that both methods, via Bessel or
(generalized) Jacobi theta functions, give the same results.
In addition the derivatives w.r.t. $p^2$ for all the integrals
are compared with taking a numerical derivative.

Note that the sunset functions at finite volume call the tadpole integrals
evaluated with the same method. Do not forget to set precision for those
as well.
The precision for the sunset integrals can be set with the
functions  
\texttt{setprecisionfinitevolumesunsett(racc,rsacc,printout)}
and \texttt{setprecisionfinitevolumesunsetb(maxsum1,mxsum2,racc,rsacc,printout)}.
The bool variable \texttt{printout} defaults to \texttt{true}
and sets whether the setting is
printed. The double values \texttt{sunsetracc} and \texttt{sunsetrsacc}
set the accuracies
of the numerical integration needed when one or two loop-momenta ``feel''
the finite volume. Default values are \texttt{1e-5} and \texttt{1e-4}
respectively.
The integers \texttt{maxsum1} and \texttt{maxsum2} give how far the sum over
Bessel functions
is used for the same two cases. The first is maximum 400,
the second maximum 40 in the present implementation.
In the latter case a triple sum is needed, hence the much lower upper bound.
For most applications it makes sense to have a higher precision for
the case with one loop momentum quantized, i.e. \texttt{racc} smaller than
\texttt{rsacc}.

\section{Masses, decay constants and vacuum-expectation-values}
\label{massdecayvev}

\subsection{Masses}

The masses of the pion, kaon and eta at two-loops in three flavour ChPT
were calculated in \cite{Amoros:1999dp}. The pion and eta mass were done earlier
with a different subtraction scheme and a different way to perform the
sunset integrals in \cite{Golowich:1997zs}.

The expressions for the physical masses for $a=\pi,K,\eta$ are given by
\begin{equation}
m_{a\,\mathrm{phys}}^2 = m_{a\,0}^2+m_a^{2(4)}+m_a^{2(6)}\,.
\end{equation}
The superscripts indicate the order of the diagrams in $p$ that each
contribution comes from. 
The lowest order masses are given in (\ref{lomasses}).
The expressions can be found in \cite{Amoros:1999dp}.
In addition the contributions themselves are split in the parts
depending on the NLO LECs $L_i^r$, on the NNLO LECs $C_i^r$ and the
remainder as
\begin{equation}
\label{defmasses}
m_a^{2(4)}=m_{a\,L}^{2(4)}+m_{a\,R}^{2(4)}\,,
\qquad
m_a^{2(6)}=m_{a\,L}^{2(6)}+m_{a\,C}^{2(6)}+m_{a\,R}^{2(6)}\,.
\end{equation}
All the parts in (\ref{defmasses}) are implemented as the functions
\texttt{mpi4(physmass,Li)},\\ \texttt{mpi4L(physmass,Li)}, 
\texttt{mpi4R(physmass)}, \texttt{mpi6(physmass,Li,Ci)},
\texttt{mpi6L(physmass,Li)}, \texttt{mpi6C(physmass,Ci)}
and \texttt{mpi6R(physmass)}\,. The equivalent functions also exist 
for the kaon, with \texttt{pi} to \texttt{k}, and eta,
with \texttt{pi} to \texttt{eta}.

The functions are defined in \texttt{massesdecayvev.h} and
implemented in \texttt{massesdecayvev.cc}.

\subsection{Decay constants}

The decay constants of the pion, kaon and eta at two-loops in three flavour ChPT
were calculated in \cite{Amoros:1999dp}. The pion and eta decay constants
were done earlier
with a different subtraction scheme and a different way to perform the
sunset integrals in \cite{Golowich:1997zs}.

The expressions for the decay constants for $a=\pi,K,\eta$ are given by
\begin{equation}
\label{defdecayfirst}
F_{a\,\mathrm{phys}} = F_0\left(1+F_a^{(4)}+F_a^{(6)}\right)\,.
\end{equation}
The superscripts indicate the order of the diagrams in $p$ that each
contribution comes from. $F_0$ denotes the decay constant in the three-flavour
chiral limit.
The expressions were originally derived
in \cite{Amoros:1999dp},
but note the description in the erratum of \cite{Amoros:2000mc}.
The expressions corrected for the error can be found in the website
\cite{webpage}.
In addition the contributions themselves are split in the parts
depending on the NLO LECs $L_i^r$, on the NNLO LECs $C_i^r$ and the
remainder as
\begin{equation}
\label{defdecay}
F_a^{(4)}=F_{a\,L}^{(4)}+F_{a\,R}^{(4)}\,,
\qquad
F_a^{(6)}=F_{a\,L}^{(6)}+F_{a\,C}^{(6)}+F_{a\,R}^{(6)}\,.
\end{equation}
All the parts in (\ref{defdecay}) are implemented as the functions
\texttt{fpi4(physmass,Li)},\\ \texttt{fpi4L(physmass,Li)}, 
\texttt{fpi4R(physmass)}, \texttt{fpi6(physmass,Li,Ci)},
\texttt{fpi6L(physmass,Li)}, \texttt{fpi6C(physmass,Ci)}
and \texttt{fpi6R(physmass)}\,. The equivalent functions also exist 
for the kaon, with \texttt{pi} to \texttt{k}, and eta,
with \texttt{pi} to \texttt{eta}. For the $\eta$ the decay constant has been
defined with the octet axial-vector current.

The functions are defined in \texttt{massesdecayvev.h} and
implemented in \texttt{massesdecayvev.cc}.

\subsection{\texttt{getfpimeta}}

A problem that occurs in trying to compare to lattice QCD is that
the present routines are written in terms of the physical
pion decay constant and masses. In particular, the eta mass is treated as
physical. One thus needs a consistent eta mass and pion decay constant when
varying the input pion and kaon mass. This assumes we have fitted the
LECs $L_i^r$ and $C_i^r$ with a known set of $m_\pi,m_K,m_\eta,F_\pi$.

The functions \texttt{getfpimeta6(mpiin,mkin,massin,Li,Ci)}
and\\ \texttt{getfpimeta4(mpiin,mkin,massin,Li)} return a \texttt{physmass}
with a consistent set of $F_\pi$ and $m_\eta$ for input values
of the pion and kaon mass. The other input is the \texttt{physmass}
\texttt{massin}, the \texttt{Li} and \texttt{Ci} that are used as input.
The formulas used are (\ref{defmasses}) and (\ref{defdecay}) up to order
$p^6$ and $p^4$ respectively.
The solution is obtained by iteration and stops when six digits of precision
are reached. This method was used in \cite{Bijnens:2014dea} to obtain the
consistent set of masses and decay constants used there.

\subsection{Vacuum-expectation-values}

The corrections to the vacuum expectation values (vevs)
$\langle0\vert \overline q q \vert 0\rangle$
for up, down and strange quarks in the isospin limit
were  calculated at two-loops in three flavour ChPT
in \cite{Amoros:2000mc}.
The expression for the up and down quark vev are identical since we
are in the isospin limit.

We write the expressions in a form analoguous to the decay constant
treatment:
\begin{equation}
\langle0\vert \overline q q \vert 0\rangle_{a\,\mathrm{phys}} =
- F_0^2 B_0\left(1+\langle0\vert \overline q q \vert 0\rangle_a^{(4)}+
\langle0\vert \overline q q \vert 0\rangle_a^{(6)}\right)\,.
\end{equation}
The superscripts indicate the order of the diagrams in $p$ that each
contribution comes from. The lowest order values are $-F_0^2 B_0$.

Note that the vevs are not directly measurable quantities. They depend
on exactly the way the scalar densities are defined in QCD. ChPT can be
used for them when a massindependent, chiral symmetry respecting subtraction
scheme is used. $\overline{MS}$ in QCD satisfies this, but there are other
possibilities. Even within a scheme, $B_0$ and the quark masses depend
on the QCD subtraction scale $\mu_\textrm{QCD}$ is such a way that
$B_0 m_q$ is independent of it. The higher order corrections in this
case also depend on the LECs for fully local counter-terms,
$H_1^r,H_2^r$ at order $p^4$ and $C_{91}^r,\ldots,C_{94}^r$ at $p^6$.
When the scalar density is fully defined, measuring these quantities in
e.g. lattice QCD and comparing with the ChPT expressions is a well defined
procedure.

The contributions at the different orders themselves are split in the parts
depending on the NLO LECs $L_i^r$, on the NNLO LECs $C_i^r$ and the
remainder as
\begin{eqnarray}
\label{defvev}
\langle0\vert \overline q q \vert 0\rangle_a^{(4)}&=&
\langle0\vert \overline q q \vert 0\rangle_{a\,L}^{(4)}
+\langle0\vert \overline q q \vert 0\rangle_{a\,R}^{(4)}\,,
\nonumber\\
\langle0\vert \overline q q \vert 0\rangle_a^{(6)}&=&
\langle0\vert \overline q q \vert 0\rangle_{a\,L}^{(6)}
+\langle0\vert \overline q q \vert 0\rangle_{a\,C}^{(6)}
+\langle0\vert \overline q q \vert 0\rangle_{a\,R}^{(6)}\,.
\end{eqnarray}
All the parts in (\ref{defvev}) are implemented as the functions
\texttt{qqup4(physmass,Li)},\\ \texttt{qqup4L(physmass,Li)}, 
\texttt{qqup4R(physmass)}, \texttt{qqup6(physmass,Li,Ci)},\\
\texttt{qqup6L(physmass,Li)}, \texttt{qqup6C(physmass,Ci)}
and \texttt{qqup6R(physmass)}\,. The equivalent functions also exist 
for the strange quark case, with \texttt{up} changed to \texttt{strange}.

The functions are defined in \texttt{massesdecayvev.h} and
implemented in \texttt{massesdecayvev.cc}.

\section{Masses and decay constants at finite volume}
\label{massdecayvevV}

The expressions treated in this section have been
derived in \cite{Bijnens:2014dea}. A general remark is that care should be
taken to set the precision in the loop integrals sufficiently high.
For the one-loop integrals setting it very high is usually no problem.
For the sunset integrals the evaluation can become very slow. It is
strongly recommended to play around with the settings and compare the outputs
for the two ways to evaluate the integral. 
The theta and Bessel function evaluation approach the correct answer
differently.
For most cases
it is possible to have \texttt{rsacc} set smaller than \texttt{racc}.

For many applications it is useful to calculate the
very time consuming parts, those labeled \texttt{6RV}, once and store them.
They only depend nontrivially on the masses and size of the finite volume.
The decay constant dependence is very simple and there is dependence
on the NLO LECs $L_i^r$.

The results presented in this section are with periodic boundary conditions
and an infinite extension in the time direction. They are also restricted
to the case where the particle is at rest, i.e. $\vec p=0$.

\subsection{Masses at finite volume}

The finite volume corrections to the masses squared\footnote{Note that
in other papers the corrections to the mass itself are sometimes
quoted.} are defined as the difference of the mass squared in finite volume
and in infinite volume:
\begin{eqnarray}
\Delta^V m^2_a &=& m^{2V}_a-m^{2\,V=\infty}_a
= m_a^{2V(4)}+ m_a^{2V(6)}\,.
\nonumber\\
m_a^{2V(6)} &=& m_{a\,L}^{2V(6)}+m_{a\,R}^{2V(6)}\,.
\end{eqnarray}
These definitions are for $a=\pi,K,\eta$.
These functions are available as
\texttt{mpi4Vt}, \texttt{mpi6Vt}, \texttt{mpi6LVt} \texttt{mpi6RVt}
respectively for the pion. \texttt{mpi4Vt} and \texttt{mpi6RVt}
have as arguments
a physmass and the length $L$. The other two have as arguments
\texttt{physmass,Li double L}. The final letter ``t'' indicates
the evaluation using  theta functions. With a ``b'' instead they use
evaluation via Bessel functions.

The equivalent functions for the kaon (pi to k) and eta (pi to eta)
are also available. All these are defined in \texttt{massdecayvevV.h}
and implemented in \texttt{massdecayvevV.h}.

\subsection{Decay constants at finite volume}

The finite volume corrections to the decay constants
 are defined as the difference of the mass squared in finite volume
and in infinite volume:
\begin{eqnarray}
\label{defdecayV}
\Delta^V F_a &=& F^{V}_a-F^{\,V=\infty}_a
= F_a^{V(4)}+ F_a^{V(6)}\,.
\nonumber\\
F_a^{V(6)} &=& F_{a\,L}^{V(6)}+F_{a\,R}^{V(6)}\,.
\end{eqnarray}
These definitions are for $a=\pi,K,\eta$.
Note that the correction is defined to the value of the
decay constant, not divided by the the lowest order decay constant as
in (\ref{defdecayfirst}).
The functions are available as
\texttt{fpi4Vt}, \texttt{fpi6Vt}, \texttt{fpi6LVt} \texttt{fpi6RVt}
respectively for the pion. \texttt{fpi4Vt} and \texttt{fpi6RVt}
have as arguments
a physmass and the length $L$. The other two have as arguments
\texttt{physmass,Li double L}. The final letter ``t'' indicates
the evaluation using  theta functions. With a ``b'' instead they use
evaluation via Bessel functions.

The equivalent functions for the kaon (pi to k) and eta (pi to eta)
are also available. All these are defined in \texttt{massdecayvevV.h}
and implemented in \texttt{massdecayvevV.h}.

\section{Various comments}
\label{various}

\subsection{Error handling}

Error handling has been dealt with in a very simple manner. Most functions
print out a message to standard output if something doesn't seem right.
In particular, since the subtraction scale is present in several
inputs, many functions check if these are the same and print out a message
if not.

Errors due to a zero in a denominator are not caught and might lead to a
crash.

\subsection{Warnings}

The definition of higher orders in ChPT is not unique. In this
program collection, we have consistently chosen to rewrite all results
in the physical masses and the physical pion decay constant, but note that
even this is not fully unique. While there is usually a standard choice for the
lowest order expression, at one-loop order this is often not the case since
the Gell-Mann--Okubo relation can be used to rewrite the dependence on the
$\eta$ mass.

The files contain many internal functions as well as some extensions
which are not described in this manuscript. These might change in future
releases and have in general not been as fully tested as the described ones.
Use at your own risk.
 
\subsection{Possible extensions}

Two-loop results are known for many more
ChPT quantities also including isospin violation as well as for
two-flavour ChPT and the partially quenched case. In addition at the one-loop
level very many extensions exist for inclusion of the internal
electro-magnetic interaction, finite volume effects, twisting and various
extensions that include finite $a$ effects in lattice gauge theory.

Another extension is how the higher orders are actually defined. In this
program collection so far we have consistently chosen to rewrite all results
in the physical masses and the physical pion decay constant. 
Implementations of other choices of higher orders, in
particular in terms of lowest-order quantities are planned.

A last extension worth mentioning is the inclusion of the existing
two-flavour ChPT results.

\section{Conclusions}
\label{conclusions}

This paper describes a library of useful numerical programs in \textsc{C++}
for ChPT at upto two-loop order. Care has been taken to be independent
of other libraries. In particular a number of numerical routines has
been reimplemented in the numerical algorithm part of the library,
\texttt{libjbnumlib.a}. The more ChPT direct functions like the
loop integrals, a number of data structures to deal with LECs and
the result for the masses, decay constants and decay constants
are put in \texttt{libchiron.a}. Finite volume results
are included for the masses and decay constants.

A simple Makefile as well as a large number of testing/example programs are
included.

\section*{Acknowledgements}

This work is supported, in part, by the European Community SP4-Capacities
``Study of Strongly Interacting Matter'' (HadronPhysics3,
Grant Agreement number 283286) and
the Swedish Research Council grants 621-2011-5080 and 621-2013-4287.
I also thank all my collaborators in the work for which my version
of the program made it into this collection and especially
Ilaria Jemos who has tested many of the earlier versions
in the course of \cite{Bijnens:2011tb}.

\end{document}